\documentclass{desyproc}

\begin{document}
\title{CMS Results on Hard Diffraction}

\author{{\slshape Christina Mesropian$^1$ on behalf of the CMS collaboration}\\[1ex]
$^1$The Rockefeller University, New York, USA}

\contribID{28}



\maketitle

\begin{abstract}
In these proceedings we present CMS results on hard diffraction. Diffractive dijet production
 in $pp$ collisions at $\sqrt{s}$=7 TeV is discussed. The cross section for dijet production 
is presented as a function of $\tilde{\xi}$, representing the fractional momentum loss of 
the scattered proton in single-diffractive events. The observation of W and Z boson production in 
 events with a large pseudo-rapidity gap is also presented.
\end{abstract}

\section{Introduction}
Diffractive processes contribute a significant fraction to the total inelastic proton-proton cross sections at high energies. 
These reactions can be described in terms of the exchange of a $pomeron$, a hypothetical object with the quantum numbers of the vacuum. 
The experimental signatures of diffractive events are the presence of non-exponentially suppressed
large rapidity gaps and/or presence of the intact leading protons. Diffractive events 
with a hard parton-parton scattering, so called {\em hard diffractive events}, subject of these proceedings, are 
of particular interest 
since they can be studied in terms of perturbative QCD.

The measurements presented here are based on the data collected by the CMS experiment during 2010 at a $\sqrt{s}=$7 TeV. 
The detailed description of the CMS experiment can be found elsewhere~\cite{CMS}. The simulation of non-diffractive events 
is  obtained with  PYTHIA6~\cite{PYTHIA6} and PYTHIA8~\cite{PYTHIA8}. Hard diffractive events are simulated with the POMPYT~\cite{POMPYT} and POMWIG~\cite{POMWIG} generators, 
as well as PYTHIA8, these generators are used with diffractive PDFs (dPDFs) from the same fit to diffractive deep inelastic 
scattering data, H1 fit B.

\section{Diffractive Dijet Production}
\begin{figure}[hb]
\centerline{
\includegraphics[width=0.5\columnwidth]{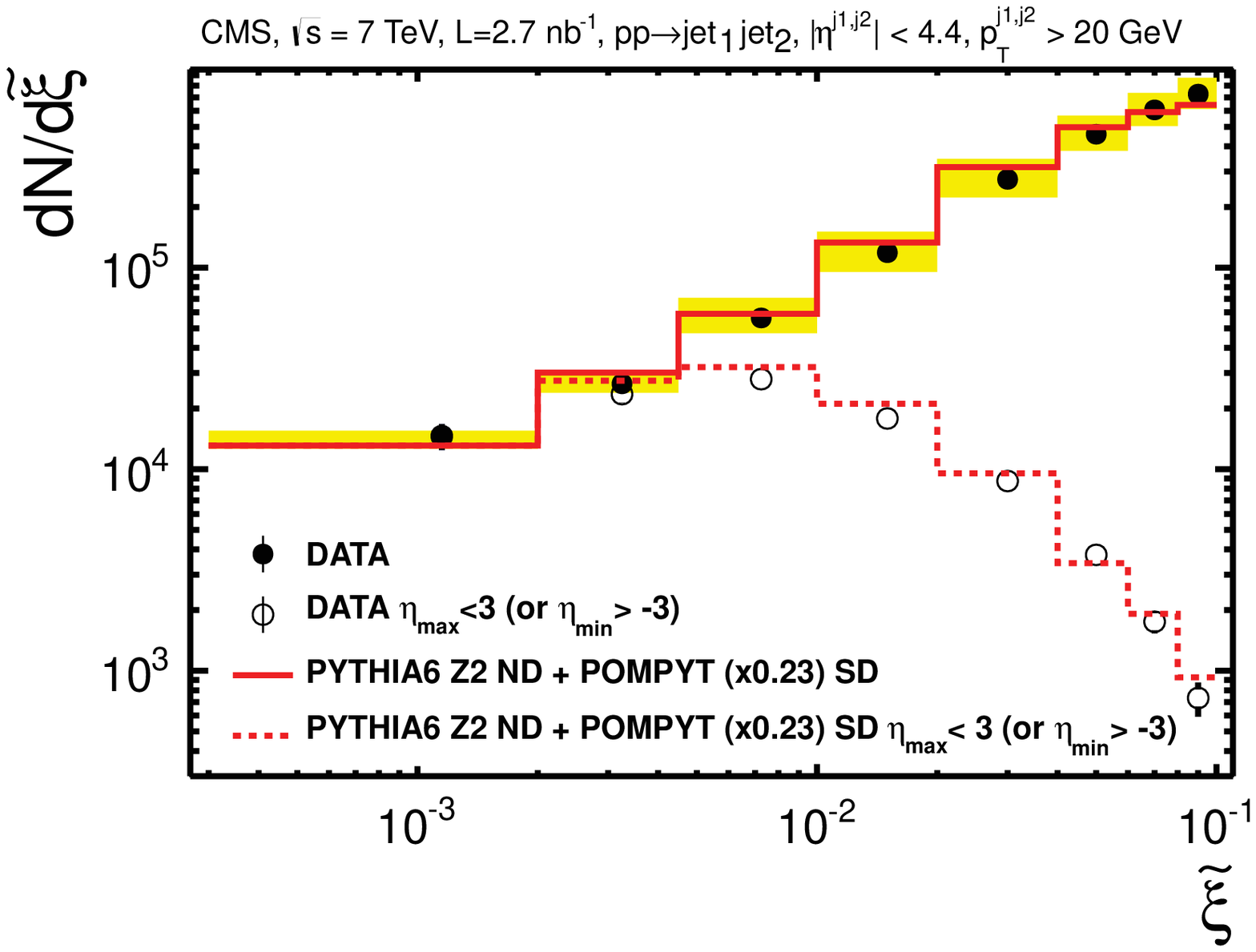}
\includegraphics[width=0.5\columnwidth]{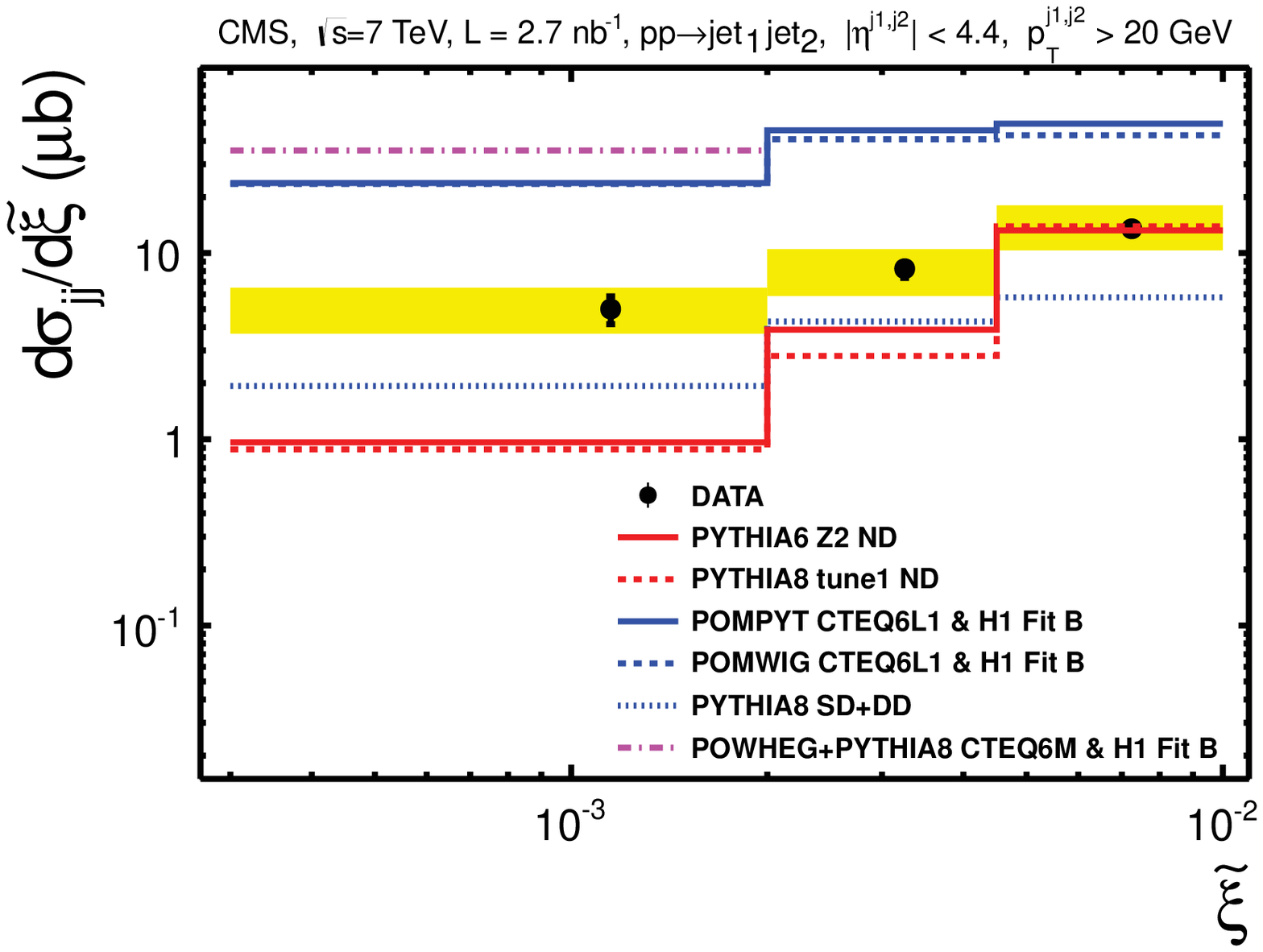}
}
\vspace{-0.5cm}
\caption{(a) Reconstructed $\tilde{\xi}$ distributions with (open symbols) and without (closed symbols) 
the $\eta_{max}<$ 3 requirement, compared to MC predictions; (b) the differential cross section for 
inclusive dijet production as a function of $\tilde{\xi}$.}
\label{diff-dijet}
\end{figure}

Dijet events were selected by requiring 2 jets with $p_T>$20 GeV and -4.4$<\eta^{j1,j2}<$4.4. The anti-$k_T$ clustering 
algorithm with the radius parameter $R$=0.5 was used to reconstruct jets. 
To enhance the diffractive contribution, the requirements $\eta_{max}<$3 ($\eta_{min}>$-3) were enforced, where $\eta_{max(min)}$ 
is the pseudorapidity of the most forward (backward) particle-flow (PF)~\cite{PF} object, combining measurement from the  tracker and 
the calorimeters. This selection translates to imposing a rapidity gap of at least 1.9 units in the Hadron Calorimeter (HF) (with the coverage 3.0$<\mid\eta\mid<$5.0).
Figure \ref{diff-dijet}(a) shows the effect of this selection, the $\eta_{max}$ requirement rejects 
events at high $\tilde{\xi}$ values, whereas region of low $\tilde{\xi}$ dominated by the diffractive contribution is not affected. 
The results are compared to MC predictions where the relative diffractive contribution was scaled by a factor of 0.23 obtained by 
minimizing the difference between $\tilde{\xi}$ distributions of the data and sum of 
non-diffractive and diffractive models. The diffractive cross section as a function of $\tilde{\xi}$ 
is shown in Fig.~\ref{diff-dijet}(b). 
The low $\tilde{\xi}$ bin shows significant contribution from diffractive dijet production, 
observed for the first time at the LHC~\cite{CMS-diff-dijets}. The associated {\em rapidity gap survival} probability is estimated to be in the range from 
0.08$\pm$0.04 (NLO) to 0.12$\pm$0.05(LO) by comparing measured cross section and predictions
 from the leading order (POMPYT and POMWIG) and next-to-leading order (POWHEG)~\cite{POWHEG} diffractive generators  
based on dPDFs from the HERA experiments.

\section{Diffractive W and Z Production}
The identification of W and Z events is based on the presence of isolated electrons and muons with high transverse momentum.
\begin{figure}[hb]
\includegraphics[width=0.45\columnwidth]{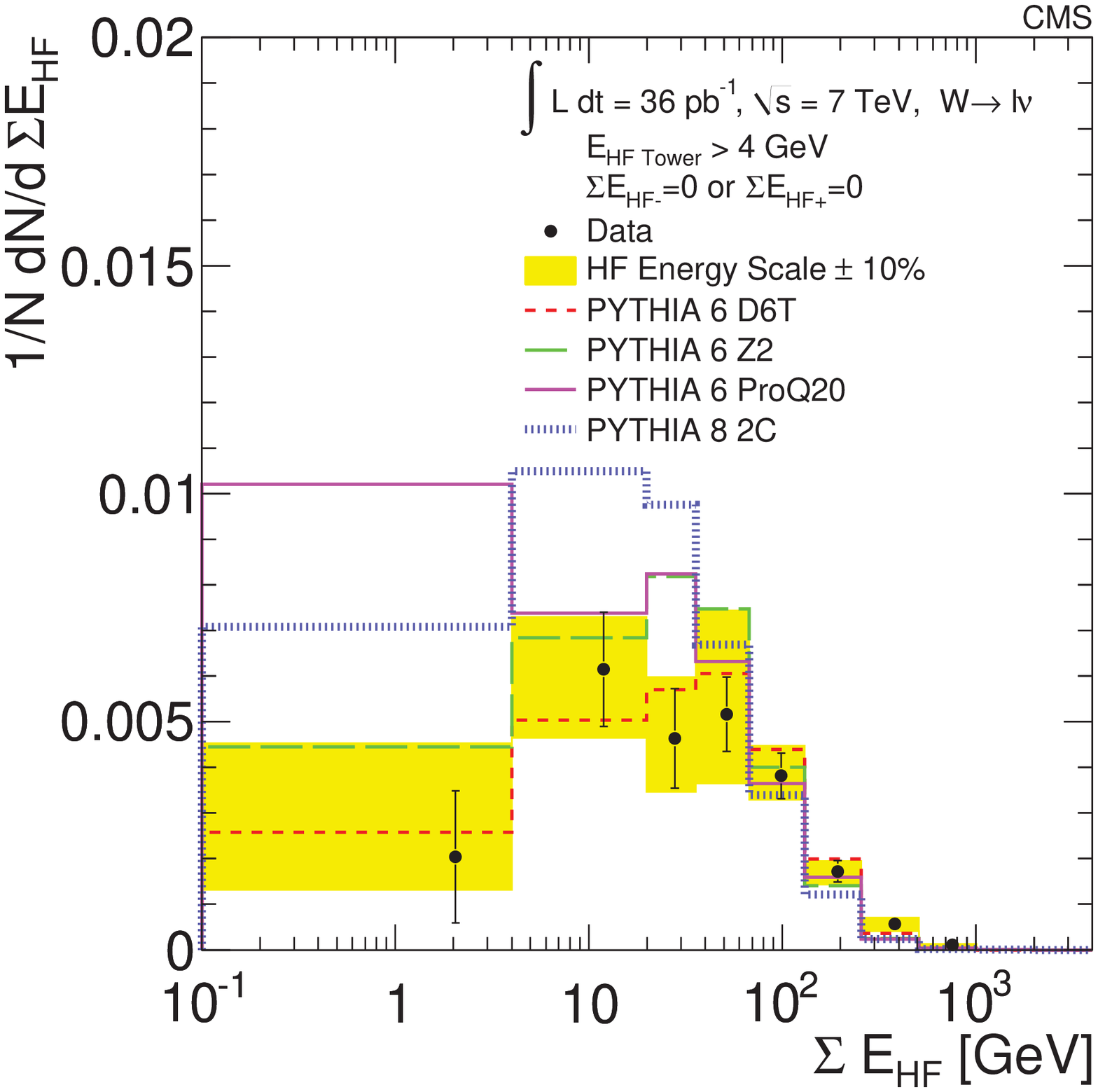}
\includegraphics[width=0.45\columnwidth]{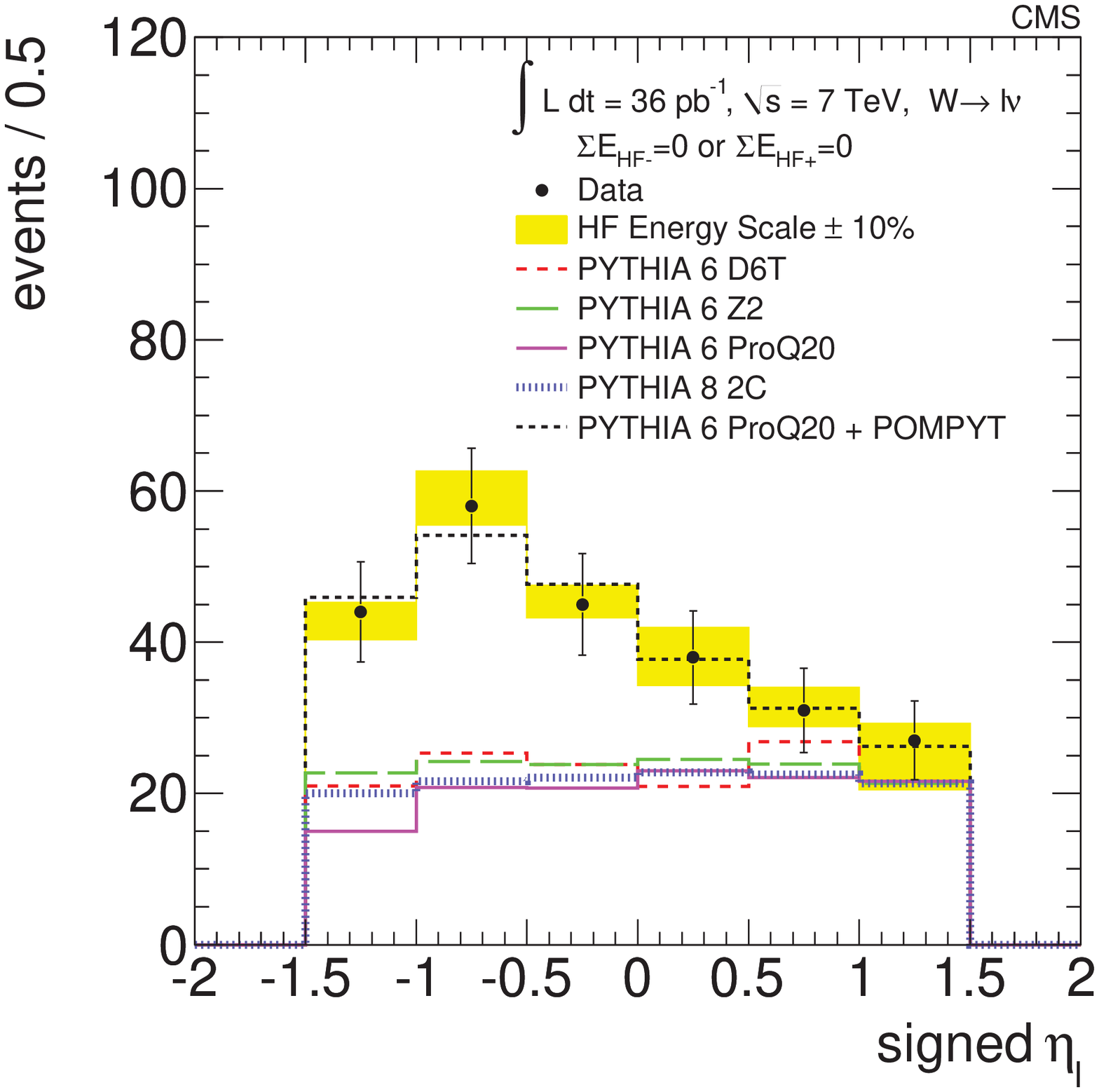}
\vspace{-0.5cm}
\caption{(a) HF energy distributions opposite to the gap in events with a large rapidity gap signature for the data and different MC tunes; (b) signed lepton pseudorapidity distribution in W events with a large rapidity gap signature, with the sign defined by the pseudorapidity of the lepton relative to the gap.}
\label{diff-W}
\end{figure}
 Electron (muon) candidates are required to have  $\mid\eta \mid<$1.4 and $p_T>$25 GeV, while events with a second isolated 
electron or muon with $p_T>$10 GeV are rejected, missing transverse momentum is required to be greater than 30 GeV and 
the transverse mass of the charged lepton and the neutrino to be greater than 60 GeV; similarly for Z candidates leptons were required to 
have $p_T>$25 GeV and $\mid\eta\mid<$1.4 with the reconstructed invariant mass of the dilepton system restricted between 60 and 120 GeV.
This selection results in essentially background-free W and Z event samples.

Fig.~\ref{diff-W}(a) shows charged particle multiplicity for W events (electron and muon channels combined) 
with  a large rapidity gap (LRG) signature, enforced by requiring that none of the calorimeter towers had 
a measured energy of more than 4 GeV in at least one of the HF calorimeters, which corresponds to a pseudorapidity  interval of 1.9 units. 
The data are compared with the predictions of PYTHIA6 and PYTHIA8. Events with zero energy deposition 
manifest the presence of a pseudorapidity gap extending over HF.  However large discrepancies between the data 
and different models are observed. 
The percentage of W and Z events with  LRG signature is (1.46$\pm$0.09(stat.)$\pm$0.38(syst.))\% and 
(1.57$\pm$0.25(stat.)$\pm$0.42(syst.))\%. 

A large asymmetry is observed between the number of events with the charged lepton 
from W decay in the opposite  and with the same hemisphere as the rapidity gap, in accordance 
with the predictions of POMPYT diffractive MC, see Fig.~\ref{diff-W}(b). In comparison, the various non-diffractive MC predictions predict a symmetric distribution, see PYTHIA curves on Fig.~\ref{diff-W}(b). The diffractive component 
is estimated to be (50$\pm$9.3(stat.)$\pm$5.2(syst.))\% by fitting the observed asymmetry with the admixture 
of POMPYT and non-diffractive (PYTHIA) events. This presents first evidence for diffractive W production at the LHC~\cite{CMS-diff-W}.

\section{Acknowledgments}

I would like to thank the organizers of 15th International Conference on Elastic \& Diffractive
Scattering for warm hospitality and for an exciting conference.


\begin{footnotesize}

\end{footnotesize}
\end{document}